\documentclass[aps,prd,twocolumn,groupedaddress,showpacs,nofootinbib]{revtex4}
\usepackage{graphicx,dcolumn,bm,amssymb,amsmath,latexsym,amsfonts,footnote,tensor}
\usepackage[usenames]{color} 
\usepackage{float}
\usepackage[colorlinks]{hyperref}
\usepackage{multirow}
\usepackage{epsfig}

\begin{document}

\title{Nonlinear electromagnetic generalization of the Kerr-Newman solution with cosmological constant}

\author{Oscar Galindo-Uriarte}
\email{oscar.galindo@cinvestav.mx}
\author{Nora Breton}
\email{nora.breton@cinvestav.mx}
\affiliation{Physics Department, \\
Centro de Investigaci\'{o}n y de Estudios Avanzados del Instituto Polit\'ecnico Nacional (Cinvestav), \\ 
PO. Box 14-740, Mexico City, Mexico }


\begin{abstract}
We present the two exact solutions of the Einstein-Nonlinear electrodynamics equations that generalize the Kerr-Newman solution. 
We determined the generalized electromagnetic potentials using  the alignment between  the tetrad vectors of the metric and the eigenvectors of the electromagnetic field tensor. It turns out that there are only two possible nonlinear electromagnetic generalizations of the Kerr-Newman geometry, corresponding to different electromagnetic potentials. The new solutions possess horizons and satisfy physical energy conditions such that they can represent  black holes with nonlinear electromagnetic charges, characterized by the parameters of mass,  angular momentum, charge, and one nonlinear parameter; the nonlinear parameter resembles the effect of a cosmological constant, negative or positive, such that the solutions are asymptotically AdS or dS. The canonical form of the electromagnetic nonlinear energy-momentum tensor is analyzed in relation with the energy conditions; it is shown that the conformal symmetry is broken by the electromagnetic nonlinear matter; the
corresponding  nonlinear electromagnetic Lagrangian as a function of the coordinates is presented as well.
\end{abstract}

\maketitle
\section{Introduction}

There are current observations of gravitational waves detecting the collision of two massive compact objects in the LIGO \cite{LIGO2016} and Virgo \cite{ALAV2021} interferometric facilities; this has lead to the assembly of catalogues of colliding compact objects that result in a unique remnant. The  astrophysical compact objects are rotating and therefore in the context of the  Einstein exact solutions there is a great interest in stationary solutions since, within some approximation, they resemble  some features of celestial bodies. Therefore the Kerr and Kerr-Newman stationary solutions of the coupled Einstein-Maxwell equations are of the utmost relevance both theoretical and astrophysically. 

On the other hand,  nonlinear electromagnetic (NLE) effects occur in the vicinity of strongly magnetized compact objects, like magnetars or neutron stars; the description of such effects require some extension of Maxwell electrodynamics and one way is with Lagrangians that are nonlinear in the electromagnetic invariants. Therefore exact solutions of the Einstein-NLE equations can give insight of interesting properties of strongly magnetized black holes (BH)  as well as can be useful as test beds of numerical simulations. Moreover,  stationary solutions  of the Einstein equations with NLE fields may open a new perspective of the physics of rotating celestial  bodies. From the theoretical point of view there are several aspects for studying stationary axisymmetric solutions that belong to the algebraic type D in the Petrov classification. An interesting possibility of introducing NLE effects in BH metrics is of avoiding the singularity, that for the static case there is abundance of regular BHs sourced by some kind of nonlinear electrodynamics \cite{AyonGarciaPRL}. However the challenge of determining a NLE stationary solution has been elusive until recently that a Euler-Heisenberg generalization of the Kerr-Newman black hole was presented in \cite{BLM2022}. 

There are several proposals of NLE Lagrangians, that are nonlinear functions of the two Lorentz invariants $F$ and $G$ of the electromagnetic field, $\mathcal{L}(F,G)$.
In this paper we emphazise that even if we do not know exactly the expression of the Lagrangian in terms of the electromagnetic invariants, new NLE solutions that generalize Kerr-Newman solution can be generated, such is the case in \cite{AGarcia2022} and \cite{AGarcia_Annals2022}, where a stationary solution of the coupled Einstein-NLE equations was presented. This exact solution is a Kerr-like geometry that describes a  rotating BH endowed with mass, angular momentum,  cosmological constant, electric charge and an electromagnetic nonlinear parameter.  The electromagnetic tensors $F_{\mu\nu}$ and $P_{\mu\nu}$  of
the solution fulfill a set of four generalized ``Maxwell equations'' and two independent Einstein--NLE
equations related with the two independent eigenvalues of the NLE
energy--momentum tensor. The NLE is determined  from a Lagrangian
that is a function of the coordinates $r$ and $\theta$, $\mathcal{L} (r, \theta)$, constructed from the two electromagnetic invariants $F$ and $G$.

There are NLE theories that are Lorentz invariant and gauge invariant, these theories were studied and classified by Pleba\'nski \cite{Pleb}, and important contributions are due to Boillat \cite{Boillat1970}.
The propagation of light in NLE environments is also of interest and
it is known that for any theory of the Pleba\'nski class the rays are the null geodesics of two optical metrics; causality and signal propagation has been addressed in \cite{Perlick2016}. The optical metric was rederived by Novello et al. \cite{Novello}; and,  using a different representation, by Obukhov and Rubilar \cite{Obukov2002}, that also derived  the Fresnel equation for the wave covectors and, for the class of local nonlinear Lagrangian nondispersive models, it is demonstrated that the quartic Fresnel equation factorizes, yielding the generic birefringence effect. 

The cosmological constant $\Lambda$ has acquired relevance lately related to its interpretation as the dark energy in cosmological solutions of the Einstein equations. Another aspect of interest are the anti–de Sitter (AdS) spacetimes ($\Lambda < 0$) related to the holographic correspondence between
gravity systems and the conformal field theory, the
AdS/CFT duality \cite{Witten98}.  Moreover, BHs in AdS spacetimes admit a gauge duality description through a thermal field theory. 
For these reasons we include the cosmological constant in our study of the nonlinear electromagnetic stationary solutions.

In this paper we present the nonlinear electromagnetic generalizations of the Kerr-Newman solutions. These new solutions are derived from aligning two vectors of the metric tetrad with the two different eigenvectors of the electromagnetic field tensor. The alignment conditions along with the condition of integrability of the Lagrangian allows to derive a differential equation for the electromagnetic potentials.  Then we consider an Ansatz
for the electromagnetic potentials $A_{\mu}$ and ${}^\star{P}_{\mu}$ that consists of a quotient of polynomials in the coordinates $r$ and $\theta$, whose coefficients are constrained by a {\it key equation}. 

The paper is organized as follows: In Sect. II we review the Kerr-Newman (KN) metric emphasizing its electromagnetic fields. In Sect. III we present the NLE equations and the alignment between two vectors of the metric tetrad with the two different eigenvectors of the electromagnetic field tensor; it is also derived the {\it key equation} that links the two electromagnetic potentials $A_{\mu}$ and ${}^\star{P}^{\mu}$. In Sect. IV  we derive the two possible NLE generalizations of the KN spacetime. 
In Sect. V we present the main features of the new NLE  generalization of the KN solution, like the horizons, ergosphere and energy conditions; the static limit that is a NLE generalization of the Reissner-Nordstrom solution is presented as well.
In Sect. VI we examine the canonical form of the NLE energy-momentum tensor $T_{\mu \nu}$ and the energy conditions satisfied by the NLE matter; it is shown that the trace of the NLE energy-momentum tensor does not vanish and then the introduction of the NLE field breaks the conformal symmetry. The expression of the NLE Lagrangian as a function of the coordinates is presented.   Finally in Sect. VII Conclusions are given.
\section{The Kerr-Newman metric}\label{sect2}

It is not exaggerated to say that the Kerr-Newman (KN) solution of the coupled Einstein-Maxwell equations is of the utmost importance, both, related to recent black hole observations as well as a theoretical object which is not completely understood in all its caveats, from thermodynamics to singularity theory, while aspects like finding interior matter that sources the exterior KN geometry are still challenges.

The Kerr-like metric with cosmological constant in Boyer-Lindquist coordinates is

\begin{equation}\label{metric1}
\begin{split}
ds^2&= \frac{a^2\sin^2{\theta} \Delta_{\theta}}{\Sigma \Xi^2}  \left( {\bf{dt}}
-\frac{a^2+r^2}{a} {\bf d \phi} \right)^2\\
&+\frac{\Sigma}{\Delta_{r}}\,{\bf dr}^2+\frac{\Sigma}{\Delta_{\theta}}\,{\bf d \theta}^2
 -{\frac{\Delta_{r}}{\Sigma \Xi^2}}\left( {\bf{dt}} -
a\sin^2{\theta}{\bf d \phi} \right)^2,
\end{split}
\end{equation}

\begin{equation}\label{solQbeta}
\begin{split}
 \Sigma(\theta,r)&:={{r^2+a^2\cos^2{\theta}}}; 
 \\ \Delta_{\theta} (\theta) &= 1 + \frac{\Lambda}{3}a^2 \cos^2 \theta \nonumber; \quad\Xi=1+\frac{\Lambda}{3}a^2,\\
 \Delta_{r}(r) &=:K(r) -2 m r+ {r}^{2}+{a}^{2}-\frac{\Lambda}{3} r^2(r^2+a^2), 
\end{split}
\end{equation}
where the angular momentum $a$, mass $m$, cosmological constant $\Lambda$ 
and  the structural function $K(r)$ is to  be determined from the coupled NLE--Einstein equations; for the Kerr-Newman solution $K(r)=Q_e^2 + Q_m^2$, with $Q_e$, $Q_m$ being the BH electric and magnetic charges, respectively;  the determinant of the Kerr-like metric  is given by

\begin{equation}\label{detg}
g=- g_{rr} g_{\theta\theta} \left(g_{t\phi}^2-g_{tt}g_{\phi\phi}\right) =-\frac{ \Sigma^2 \sin^2\theta}{\Xi^4}.
\end{equation}
Due to the existence of two Killing vectors
$\partial_{t}$ and $\partial_{\phi}$, in this spacetime the energy and angular momentum of a test particle, $E$ and $L$,  are conserved quantities, besides its mass, $\mu$.  Moreover,  the existence of a Killing tensor (additional to $g_{\mu \nu}$) implies a fourth motion constant, the Carter {\it constant} given by

\begin{eqnarray}
 C&=& \Delta_\theta P^2_\theta+ \cos^2\theta \left[a^2\left(\mu^2-\frac{\Xi E^2}{\Delta_\theta}\right)  
 +\frac{\Xi^2 L^2}{\Delta_\theta \sin^2\theta}\right] \nonumber\\
&& +\frac{\Xi^2(aE-L)^2}{\Delta_\theta},
\end{eqnarray}
where $P_{\theta}$ is the $\theta$- component of the test particle momentum. This property of the Kerr-like spacetimes  allows the separability of the Hamilton-Jacobi and the Klein-Gordon equations \cite{Frolov2007}, among other remarkable features.


The alignment  can be understood using the null tetrad $\bf{e^a}$ for the Kerr-like metric given by

\begin{displaymath}
 \left.\begin{array}{cc}
 {\bf{e^{1}}}\\
 {\bf{e^{2}}}
\end{array}\right\}
=  \sqrt{\frac{\Sigma}{2\Delta_{\theta}}} \,{\bf d \theta}\pm
i\,\frac{a\sin{\theta} }{\Xi}\sqrt{\frac{\Delta_{\theta}}{2\Sigma}}\left( {\bf{dt}}
-\frac{a^2+r^2}{a} {\bf d \phi} \right),
\end{displaymath}

and

\begin{displaymath}
 \left.\begin{array}{cc}{
 \bf {e^{3}}}\\  {\bf{e^{4}}}
\end{array}\right\}
=\sqrt{\frac{\Delta_{r}}{2\Sigma \Xi^2}} \left({\bf{dt}}-a\sin^2{\theta}{\bf d \phi} \right)\pm\, \sqrt{\frac{\Sigma} {2\Delta_{r}}}\,{\bf d r},
\end{displaymath}

the metric in the null tetrad is written as
\begin{equation}\label{null.Tetrad}
g=2{\bf{e^1}\bf{e^2}}-2{\bf{e^3}\bf{e^4}}=g_{\bf{a}\bf{b}}{\bf{e^a}\bf{e^b}},\quad
\bf{e^a}= \bf{e^a} {}_{\mu} \,\bf{dx}^{\mu}.
\end{equation}

The null tetrad is associated to the eigenvector basis of $F_{\mu\nu}$ as
\begin{equation}
F_{\mu\nu}= 2  F_{\bf{a}\bf{b}}{\bf{e^a}_{[\mu } \bf{e^b}_{\nu ] }} = 2F_{12}{e^1}_{[\mu
}\,{e^2}_{\nu]}+2F_{34}{e^3}_{[\mu }\,{e^4}_{\nu]}.
\end{equation}

In the null tetrad the EM equations can be written in terms of a closed 2--form $\omega$, as $d\omega=0$, with $\omega$ given by
\begin{equation}
    \begin{split}
 \omega&=\frac{1}{2}\left(F_{\mu\nu}+{{}^\star
P_{\mu\nu}}\right)dx^\mu\,\wedge
dx^\nu\\
&=\frac{1}{2}\left(F_{\bf{a} \bf{b}}+{{}^\star P_{\bf{a} \bf{b}}}\right)e^{\bf{a}}
\wedge\,e^{\bf{b}}\\
&= (F_{12}+ P_{34})e^{1} \wedge e^{2}+  (F_{34}+ P_{12})e^{3} \wedge e^{4}.       
    \end{split}
\end{equation}

Throughout the paper we shall mainly use the coordinate components of the electromagnetic quantities, $F_{\mu \nu}(r, \theta)$, etc.

\subsection{Alignment conditions for the  Kerr-like metric}

The field tensor $F_{\mu\nu}$ is characterized by four nonvanishing components: $
F_{\theta\phi}$,
 $F_{\theta\,t}$, $ F_{r\phi}$, $F_{rt} $. The eigenvectors $V_a^\mu$ of the tensor $F_{\mu\nu}$ are determined by solving the corresponding  eigenvalue problem; the alignment of the eigenvectors $V_a^{\mu}$, $a =1\cdots4$ along the tetrad basis, or equivalently, aligning the tetrad along the eigenvectors $V_a^{\mu}$ gives rise to the alignment conditions:
 \begin{equation}
     \begin{split}
         \label{AlignF}
{F_{r\phi}} & =  -a \, \sin^{2} { \theta }{ F_{rt}},\\
{ F_{\theta t}} & = -{\frac {{a}^{2}+{r}^{2}}{a}}{ F_{\theta \phi}},
     \end{split}
 \end{equation}

thus  only two of the field components are independent, say  $F_{rt}$ and $F_{\theta t}$, while the remaining two ${F_{r\phi}} $ and
$F_{\theta\phi}$ are determined through the alignment conditions.
Since the field tensor ${F_{\mu\nu}}$ is a curl, it can be
determined from a vector potential
$A_\mu$, $F_{\mu\nu}=A_{\nu,\mu}-A_{\mu,\nu} $. The alignment
conditions can be integrated for the electromagnetic vector
components $A_t$ and $A_{\phi}$: replacing
$F_{\theta\phi}=A_{\phi,\theta}$, $F_{r\phi}=A_{\phi,r}$, $F_{r
t}=A_{t,r}$, $F_{\theta t}=A_{t,\theta}$ in the alignment conditions Eq. (\ref{AlignF}) one arrives at

 \begin{equation}
 \begin{split}
     \label{SysAphi}
A_{\phi,r}+a \,\sin^{2} {\theta} A_{t,r} & =  0,\\
A_{\phi,\theta}+{\frac {{a}^{2}+{r}^{2}}{a}} A_{t,\theta} & =  0;
 \end{split}
 \end{equation}
while the integrability of $A_\phi$, $A_{\phi , r \theta}= A_{\phi ,  \theta r}$ leads to a partial differential equation for $A_t$
\begin{equation}\label{potAt0}
 A_{t,\theta r}
  + \frac{2 r}{\Sigma} A_{t,\theta} -\frac{2 {a}^{2}\sin {\theta} \cos {\theta}}{\Sigma} A_{t,r} =0,
\end{equation}
whose general solution has the form 

\begin{equation}\label{GS.At}
{ A_t} ={\frac {X( r ) + Y ({\theta}) }{\Sigma}},
 \end{equation}
where $X( r )$ and $Y(\theta)$ are arbitrary functions on their respective variables.
This solution for $A_t$ guarantees the  integrability  of
$A_{\phi}$, in Eq.(\ref{SysAphi}), that leads to 
\begin{equation}\label{potAphi0}
A_\phi =- a \sin^{2} {\theta } \frac{X ( r)}{\Sigma}
-  \frac{ \left({a}^{2}+{r}^{2} \right)}{a} \frac{ Y ({\theta})}{ \Sigma }.
\end{equation}

For the KN solution the electromagnetic potentials are given by the simplest polynomials,
$X(r)=Q_e\, r$ and $Y(\theta)=Q_m\, a \cos{\theta}$, where $Q_e$ and $Q_m$ are constants related to the electric and magnetic charges respectively. 
Deriving $A_\mu$ we obtain the electromagnetic field components,

\begin{equation}\label{Fmunu.KN}
    \begin{split}
        F_{r t}^{KN} &=  - \frac{1}{\Sigma^2} \left[2 Q_m a r \cos {\theta} + Q_e (r^2- a^2 \cos{\theta}^2)  \right],\\
F_{\theta t}^{KN} &=  \frac{a \sin {\theta}}{\Sigma^2} \left[ 2 Q_e\,a r \cos {\theta} - Q_m (r^2- a^2 \cos{\theta}^2)  \right],
    \end{split}
\end{equation}

Comparing with the Kerr-Newman solution we identify $Q_e$ and $Q_m$ with the electric and magnetic BH charges, respectively. 
\section{Nonlinear electromagnetic equations}

If we consider a Lagrangian, $\mathcal{L}(F,G)$, 
that depends in general form on the electromagnetic invariants $F$ and $G$, given by

\begin{align}
F &= \frac{1}{4} F_{\mu \nu} F^{\mu \nu} = \frac{1}{2} \left( B^{2}- E^{2} \right), \\
G&= \frac{1}{4} F_{\mu \nu} {}^\star{F}^{\mu \nu} = - \vec{E} \cdot \vec{B},
\end{align}
where  ${}^\star{F}_{\mu \nu} = \frac{1}{2}\sqrt{-g}\varepsilon_{\mu \nu \alpha\beta}F^{\alpha \beta}$ is the dual stress-tensor; the contravariant dual ${}^{\star}F^{\mu\nu}=-\frac{1}{2}\frac{\varepsilon^{\mu \nu \alpha\beta}}{\sqrt{-g}}F_{\alpha \beta}$;  and 
$\varepsilon^{\mu \nu \alpha\beta}$ is the Levi-Civita symbol.

The variation of the action with minimal coupling between $\mathcal{L}(F,G)$ and Einstein General Relativity leads us to define a new skew symmetric tensor  $P_{\mu \nu}$ in terms of $F_{\mu \nu}$ and the derivatives of the Lagrangian respect to the invariants, this equation is called the constitutive or material equation,

\begin{equation}\label{ConstEqP}
P_{\mu\nu}=  \mathcal{L}_F F_{\mu\nu}+ \mathcal{L}_{{G}} ({}^{\star}{F}_{\mu\nu}),
\end{equation}
where $\mathcal{L}_X =\frac{\partial \mathcal{L}}{\partial X}$; in terms of $P_{\mu\nu}$ and  $F_{\mu\nu}$ the electromagnetic field (EM) equations are
\begin{eqnarray}
{P^{\mu\nu}}_{;\nu}=0
\rightarrow{\left[\sqrt{-g}\left(\mathcal{L}_F \,{F^{\mu\nu}}+ \mathcal{L}_G \,{{}^\star{F^{\mu\nu}}}\right)\right]_{,\nu}=0},\label{Pmunu}\\
{{}^\star{F^{\mu\nu}}}_{;\nu}=0 \rightarrow {(\sqrt{-g}{}^\star {F^{\mu\nu}})_{,\nu}=0 },
\end{eqnarray}
where $g$ is the determinant of the  metric. 
Since $F_{\mu\nu}$ and ${{}^\star P_{\mu\nu}}$ are curls, both can be derived from two electromagnetic potentials, $A_{\mu}$ and ${{}^\star P_{\mu}}$,
namely,  $F_{\mu\nu} = A_{\nu,\mu}-  A_{\mu,\nu}$ and ${{}^\star P_{\mu\nu}} ={}^{\star} P_{\nu,\mu}-{}^{\star} P_{\mu,\nu}$. The nonvanishing components of the 
dual tensor ${}^{\ast}P_{\mu \nu}$  in the Kerr-like geometry (\ref{metric1}) are  $ {}^{\ast}P_{\theta t}={}^{\ast}P_{t, \theta}$ and  $ {}^{\ast}P_{r t}={}^{\ast}P_{t, r}$. Then the  EM field equations (\ref{Pmunu}) become

\begin{equation}
\left(  \sqrt{-g} P^{\phi \theta} \right)_{, \theta}  + \left(   \sqrt{-g} P^{\phi r} \right)_{, r}=0.
\end{equation}

The material equations (\ref{ConstEqP}) and the fact that ${}^{\star}P_{t,r\theta}={}^{\star}P_{t,\theta r}$ allows to write them in matrix form as,

\begin{equation}
\begin{pmatrix}
\frac{F_{\theta t}}{a\sin\theta} & -F_{rt}\\
a\sin\theta F_{rt} &F_{\theta t}
\end{pmatrix}
\begin{pmatrix}
\mathcal{L}_F\\
\mathcal{L}_G
\end{pmatrix}
=
\begin{pmatrix}
{}^{\star}P_{t,r}\\
-{}^{\star}P_{t,\theta}
\end{pmatrix},
\end{equation}
from which we clear out the derivatives of the Lagrangian,
\begin{equation}\label{LF.LG}
\begin{pmatrix}
\mathcal{L}_F\\
\mathcal{L}_G
\end{pmatrix}
=\frac{1}{\left(\frac{F_{\theta t}}{a\sin\theta}\right)^2+\left(F_{rt}\right)^2}
\begin{pmatrix}
\frac{F_{\theta t}}{a\sin\theta} & \frac{F_{rt}}{a\sin\theta}\\
- F_{rt} &\frac{F_{\theta t}}{a^2\sin^2\theta}
\end{pmatrix}
\begin{pmatrix}
{}^{\star}P_{t,r}\\
-{}^{\star}P_{t,\theta}
\end{pmatrix}.
\end{equation}



For $\mu=t$ in (\ref{Pmunu}), $\left(   \sqrt{-g} P^{t r} \right)_{, r}  + \left(   \sqrt{-g} P^{t \theta} \right)_{, \theta}=0$, we obtain the relations
\begin{align}
\sqrt{-g} P^{t r}&=\frac{\partial {}^{\star }P_\phi}{\partial \theta}=\left(\frac{r^2+a^2}{a}\right)\left(\mathcal{L}_F a\sin\theta \frac{\partial A_t}{\partial r}+\mathcal{L}_G\frac{\partial A_t}{\partial \theta}\right),\label{ptr} \\ 
\sqrt{-g} P^{t \theta}&=-\frac{\partial {}^{\star }P_\phi}{\partial r}= a\sin^2\theta\left(\frac{\mathcal{L}_F}{a\sin\theta}\frac{\partial A_t}{\partial \theta }-\mathcal{L}_G \frac{\partial A_t}{\partial r}\right)\label{pttheta}.  
\end{align}
Similarly for $\mu=\phi$ in (\ref{Pmunu}), $\left(   \sqrt{-g} P^{\phi r} \right)_{, r}  + \left(   \sqrt{-g} P^{\phi \theta} \right)_{, \theta}=0$,
\begin{align}
\sqrt{-g} P^{\phi r}&=-\frac{\partial {}^{\star }P_t}{\partial \theta}=\mathcal{L}_F a\sin\theta \frac{\partial A_t}{\partial r}+\mathcal{L}_G\frac{\partial A_t}{\partial \theta} , \label{pphir}\\ 
\sqrt{-g} P^{\phi \theta}&=\frac{\partial {}^{\star }P_t}{\partial r}= \frac{\mathcal{L}_F}{a\sin\theta}\frac{\partial A_t}{\partial \theta }-\mathcal{L}_G \frac{\partial A_t}{\partial r} .\label{pphitheta}
\end{align}
Comparing Eqs. (\ref{ptr}) - (\ref{pphitheta}) we arrive at the alignment conditions
\begin{eqnarray}\label{AlignmP}
{{}^{\star}P_{\phi,r}} & = & -a \, \sin^{2} { \theta }
 { {}^{\star}P_{t,r}}, \nonumber\\
{ {}^{\star}P_{\phi,\theta}} & = & -{\frac {
   {a}^{2}+{r}^{2}  }{a}}{ {}^{\star}P_{t,\theta}},
\end{eqnarray}
that are similar to the ones for $F_{\mu\nu}$ in Eqs. (\ref{AlignF}). Solving the system we arrive at the following solution for the vector potential ${}^{\star} P_{\mu}$,
\begin{equation}\label{GS.Pt}
{ {}^{\star} P_t} = \frac {A(r) + B({\theta})}{\Sigma}, 
 \end{equation}

\begin{equation}\label{potPphi0}
{}^{\star} P_\phi =-{a \sin^{2}{\theta }\frac { A( r )}{\Sigma}} - \left(\frac { {a}^{2}+{r}^{2} }{a} \right) \frac{B ({\theta})}{ \Sigma},
\end{equation}
where $A( r )$ and $B(\theta)$ are arbitrary functions on their respective variables.

Therefore we have derived the alignment conditions for  $P_{\mu\nu}$ in a Kerr-like metric, showing then that the alignment conditions are independent of the kind of electrodynamics under study.

For the KN solution the ${}^{\star}P_\mu$ potential is given by the polynomials,
$A(r)=Q_m\, r$ and $B(\theta)=-Q_e\, a \cos{\theta}$. Deriving ${}^{\star}P_\mu$ we obtain the dual field components of ${}^{\star}P_{\mu\nu}$,

\begin{equation}\label{Pmunu.KN}
    \begin{split}
        {}^{\star}P_{r t}^{KN} &=   \frac{1}{\Sigma^2} \left[2 Q_e a r \cos {\theta} - Q_m (r^2- a^2 \cos{\theta}^2)  \right],\\
{}^{\star}P_{\theta t}^{KN} &=  \frac{a \sin {\theta}}{\Sigma^2} \left[ 2 Q_m\,a r \cos {\theta} + Q_e (r^2- a^2 \cos{\theta}^2)  \right].
    \end{split}
\end{equation}

We can found the next relations between the components of $F_{\mu\nu}$ and ${}^{\star}P_{\mu\nu}$

\begin{equation}
    \begin{split}
        {}^{\star}P_{\theta t}^{KN}&=-a\sin\theta F_{rt}^{KN}\\
        F_{\theta t}^{KN}&=a\sin\theta \,{}^{\star}P_{rt}^{KN}.
    \end{split}
\end{equation}

Some expressions that we shall use later are  the invariants in terms of the potential $A_{t}$,
\begin{equation}
    \begin{split}
    \label{invFG}
F&= \frac{\Xi^2}{2} \left[ \left( \frac{F_{\theta t}}{a \sin  \theta} \right)^2 - \left( F_{rt} \right)^2  \right]\\
& = \frac{\Xi^2}{2} \left[ \left( A_{t,x} \right)^2 - \left( A_{t,r} \right)^2  \right],
    \end{split}
\end{equation}

\begin{equation}
    \begin{split}
    G&= -  \Xi^2\left(  \frac{F_{\theta t}}{a \sin  \theta}  F_{rt}  \right)\\
    &=\Xi^2 \left( A_{t,x} \right) \left( A_{t,r} \right),\label{FG_At}
    \end{split}
\end{equation}

\begin{equation}
    \begin{split}
\sqrt{F^2+G^2}&= \frac{\Xi^2}{2} \left[ \left( \frac{F_{\theta t}}{a \sin  \theta} \right)^2 + \left( F_{rt} \right)^2  \right]\\
&= \frac{\Xi^2}{2} \left[ \left( A_{t,x} \right)^2 + \left( A_{t,r} \right)^2  \right],
    \end{split}
\end{equation}

where we have written $F$ and $G$  in terms of the coordinate $x= a \cos \theta$, $dx = - a \sin \theta$. And the material or constitutive Eqs. (\ref{ConstEqP}) are equivalent to:

\begin{eqnarray}\label{ConstEqPx}
{}^{\ast}P_{t, r}&=& - F_{rt}  \mathcal{L}_{G}-F_{xt}  \mathcal{L}_{F}, \nonumber\\
{}^{\ast}P_{t, x}&=&  - F_{rt}  \mathcal{L}_{F} - F_{xt}  \mathcal{L}_{G}.
\end{eqnarray}
\subsection{The key equation for the NLE potentials $A_{\mu}$ and ${}^{\star} P_{\mu}$}

The electromagnetic potentials  $A_{\mu}$ and ${}^{\star} P_{\mu}$ are not independent of each other but they are constrained 
by the constitutive Eq. (\ref{ConstEqP}) that links $P_{\mu\nu}$ and $F_{\mu\nu}$. Moreover they must  be in agreement with the integrability condition or closure condition of the Lagrangian, $d^2  \mathcal{L}=0$, 
that amounts to
\begin{equation}\label{closureL}
\frac{\partial^2  \mathcal{L}}{\partial {\theta}  \partial r}= \frac{\partial^2  \mathcal{L}}{\partial {r}  \partial {\theta}}.
\end{equation}

To determine the derivatives of the Lagrangian with respect to the coordinates $r$ and $\theta$ we use  the chain rule,
\begin{equation}
 \frac{\partial  \mathcal{L}}{\partial x^{\alpha}} =\mathcal{L}_F \frac{\partial F}{\partial x^{\alpha}}+ \mathcal{L}_G \frac{\partial G}{\partial x^{\alpha}},   
\end{equation}
and the relations (\ref{LF.LG}) and (\ref{invFG}) to obtain 
\begin{align}
    \frac{\partial  \mathcal{L}}{\partial r}&=\frac{\Xi^2}{a\sin\theta}\left[\frac{\partial {}^{\star}P_t}{\partial r}\frac{\partial F_{\theta t}}{\partial r}+\frac{\partial {}^{\star}P_t}{\partial \theta}\frac{\partial F_{r t}}{\partial r}\right]\\
    \frac{\partial  \mathcal{L}}{\partial \theta}&=\frac{\Xi^2}{a\sin\theta}\left[-\frac{\cos\theta}{\sin\theta}F_{\theta t}\frac{{}^{\star}P_t}{\partial r}+\frac{\partial F_{\theta t}}{\partial \theta}\frac{\partial {}^{\star}P_t}{\partial r}+\frac{\partial F_{rt}}{\partial \theta}\frac{\partial {}^{\star}P_t}{\partial \theta}\right].
\end{align}
Replacing the fields in terms of their potentials, $F_{rt}= A_{t,r}$, $F_{\theta t}= A_{t, \theta}$, 
${}^{\star} P_{\theta t}={}^{\star} P_{t, \theta }$ and ${}^{\star} P_{r t}={}^{\star} P_{ t, r}$ in the closure condition (\ref{closureL}) we arrive at 

\begin{equation}\label{key.eq}
\left( \frac{\partial^2 A_{t}}{\partial r  \partial r} \right) \frac{\partial}{\partial {\theta}} 
\left( \frac{1}{\sin{\theta}} \frac{\partial {}^{\star} P_{t}}{\partial {\theta}}\right)-\left( \frac{\partial^2 {}^{\star} P_{t}}{\partial r  \partial r} \right)
\frac{\partial}{\partial {\theta}} 
\left( \frac{1}{\sin{\theta}} \frac{\partial A_{t}}{\partial {\theta}}\right) =0.
\end{equation}
This equation has been called {\it the key equation} in \cite{AGarcia_Annals2022}.
From the key equation we obtain constraints for the functions $A(r)$, $B(\theta)$, $X(r)$ and $Y(\theta)$ in Eqs. (\ref{GS.At}) and (\ref{GS.Pt}). 
Therefore, even if we do not know the explicit dependence of the Lagrangian with respect to the electromagnetic invariants $F$ and $G$,
we derived a key equation that must be fulfilled by the electromagnetic potentials of the electromagnetic fields in a Kerr-like spacetime.  
\section{The nonlinear electromagnetic generalizations of the Kerr-Newman metrics}

From the expressions of the electromagnetic potentials $A_t$ and ${}^{\star} P_t$ in terms of the arbitrary functions $A(r)$, $B(\theta)$, $X(r)$ and $Y(\theta)$ Eqs. (\ref{GS.At}) and (\ref{GS.Pt}), one could think at first that we can find an infinite number of NLE solutions that generalize the Kerr-Newman, just taking more terms in the polynomials $A(r)$, $B(\theta)$, $X(r)$ and $Y(\theta)$. With that in mind, we try with polynomials
for the electromagnetic potential components $A_{t}= [X(r)+ Y(\theta)]/\Sigma$ and ${}^{\star} P_{t}= [A(r)+ B(\theta)]/\Sigma$, with $X(r)$, $A(r)$, $Y(\theta)$ and $B(\theta)$ according to the Ansatz,

\begin{equation}
    \begin{split}
        X(r) & = Q_e r + \sum_{n=-5}^{30} C_{n} r^n,\\
Y (\theta) & = Q_m \,a \cos{\theta} + \sum_{s=-5}^{30} D_{s} \cos^s {\theta},\\
A(r) & = Q_m \, r + \sum_{l=-5}^{30} H_{l} r^l,\\ 
B(\theta) & = - Q_e \,a \cos{\theta} +  \sum_{k=-5}^{30} G_{k} \cos^k {\theta}, 
    \end{split}
\end{equation}
where $C_{n}$, $D_{s}$, $H_{l}$ and $G_{k}$ are constants that 
are constrained by the key equation (\ref{key.eq}). Substituting  $A_{t}$ and ${}^{\star} P_{t}$ into the key equation and equating powers it turns out that there are only two nontrivial cases of new nonlinear electromagnetic generalizations of the Kerr-Newman metric, that we have called the cubic vector potential and the quartic vector potential.
Numerically are obtained several cases with $A_t = ({\rm const}) {}^{\star}P_{t} $ that are actually trivial since they reduce to the Maxwell case. 
\subsection{Case 1. The cubic vector potential}

The electromagnetic potentials $A_{t}= [X(r)+ Y(\theta)]/\Sigma$ and ${}^{\star} P_{t}= [A(r)+ B(\theta)]/\Sigma$ of one NLE generalization of the KN solution are given by,

\begin{equation}
    \begin{split}
X(r) & = Q_e r(1-\beta r^2),\\
Y (\theta) &=  aQ_m \cos {\theta}(1+\beta a^2 \cos^2 \theta), \\
A(r) & = Q_m r(1-\beta r^2),\\ B(\theta) &=  -a Q_e \cos {\theta}(1+\beta a^2 \cos^2 \theta),
    \end{split}
\end{equation}
where $\beta$ is the nonlinear parameter. By deriving the electromagnetic potentials, we obtain the electromagnetic field components, which can be expressed as the KN field components plus the nonlinear contribution,

\begin{equation} \begin{split}
F_{rt}&= F_{rt}^{KN}- \frac{\beta r}{\Xi \Sigma^2} [2Q_m  a^3 \cos^3 {\theta} \\
&+ Q_e r(r^2+3a^2\cos^2\theta)  ],\\
F_{\theta t}&= F_{ \theta t}^{KN}- \frac{\beta }{\Xi \Sigma^2} a^2 \cos{\theta} \sin{ \theta} [ 2Q_e\,r^3 \\
&+ Q_m\, a \cos{\theta} (3 r^2 + a^2 \cos^2 {\theta}) ],\\
{}^{\star}P_{rt}&= {}^{\star}P_{rt}^{KN}- \frac{\beta r}{\Xi \Sigma^2} [- 2Q_e  a^3 \cos^3 {\theta} \\
&+ Q_m r(r^2+3a^2\cos^2\theta) ],\\
{}^{\star}P_{\theta t}&= {}^{\star}P_{ \theta t}^{KN}- \frac{\beta }{\Xi \Sigma^2} a^2 \cos{\theta} \sin{ \theta} [ 2Q_m\,r^3 \\
&- Q_e\, a \cos{\theta} (3 r^2 + a^2 \cos^2 {\theta}) ].
\end{split}
\end{equation} 
where the  KN fields $F_{\mu \nu}^{KN}$ and ${}^{\star}P_{\mu \nu}$ are given in Eqs. (\ref{Fmunu.KN}) and (\ref{Pmunu.KN}).
The metric functions $\Delta_{r}$ and $K(r)$ in the Kerr-like metric, Eq. (\ref{metric1}), are given by

\begin{eqnarray}
\Delta_{r} & = & r^2- 2mr+a^2- \frac{\Lambda}{3} r^2(r^2+a^2) + K(r),    \nonumber\\
K(r) & = & ({Q_e}^2+{Q_m}^2) (1- \beta r^2)^2,
\end{eqnarray}
and $\beta =0$ corresponds to the Kerr-Newman solution. 
The {\it cubic vector potential}  was presented in \cite{AGarcia_Annals2022} and represents a rotating nonlinearly charged BH characterized by its mass $m$, angular momentum $a$, cosmological constant $\Lambda$, electric and magnetic charges $Q_e$ and $Q_m$, and the nonlinear parameter $\beta$. Its asymptotics  can be de Sitter or anti-de Sitter, and even flatness, depending on the value  of the nonlinear parameter and of the cosmological constant. This BH can present one, two or three horizons. the third one being the cosmological  horizon in the de Sitter case.  Among the main differences between the NLE solution and  the KN-BH  are  the equatorial asymmetry that is enhanced by the NLE field and for charged particles the access to one of the poles is forbidden; besides,  a second circular orbit in the neighborhood of the external horizon appears; the presence of the nonlinear electromagnetic field increases the curvature producing bounded orbits closer to the horizon more details can be found in \cite{Breton2022}. In case the BH is static ($a=0$), the solution corresponds to a NLE generalization of the Reissner-Nordstrom solution with cosmological constant.
\subsection{Case 2.  The quartic vector potential}

The second NLE generalization of the KN solution is characterized by the electromagnetic potentials,

\begin{equation}
    \begin{split}
        {\Xi \Sigma} A_t &= Q_e\,r +Q_m\,a\cos \theta -\frac{\xi Q_e}{4} \left( r^4 + a^4 \cos^4 \theta \right), \\
{\Xi \Sigma} A_\phi&= -a\sin^2\theta Q_e\,r\left(1-\xi\frac{r^3}{4}\right)\\
&-(r^2+a^2)\cos\theta\left(Q_m-\xi\frac{Q_ea^3\cos^\theta}{4}\right),\\
 {\Xi \Sigma}  {}^{\star} P_t&= Q_m\,r - Q_e\, a\cos \theta -\frac{\xi Q_m}{4} \left( r^4 + a^4 \cos^4 \theta \right), \\
{\Xi \Sigma} {}^{\star} P_\phi&=-a\sin^2\theta Q_m\,r\left(1-\xi\frac{r^3}{4}\right)\\
&+(r^2+a^2)\cos\theta\left(Q_e+\xi\frac{Q_m \,a^3\cos^\theta}{4}\right),
    \end{split}
\end{equation}
where $\xi$ is the nonlinear electromagnetic parameter.
These potentials generate the fields
$F_{\mu\nu}=\partial_\mu A_\nu-\partial_\nu A_\mu$ and 
${}^{\star} P_{\mu\nu}=\partial_\mu( {}^{\star} P_\nu)-\partial_\nu ({}^{\star}P_\mu)$, which are explicitly written as the KN field components plus the nonlinear contribution,

\begin{equation} \begin{split}
F_{rt}^{NLE}&= F_{rt}^{KN}+\xi\frac{Q_e \,r}{2\Xi} \\
F_{\theta t}^{NLE}&= F_{ \theta t}^{KN}+\xi\frac{a^2 Q_e\,\sin\theta\cos\theta}{2\Xi},\\
{}^{\star}P_{rt}^{NLE}&= {}^{\star}P_{rt}^{KN}+\xi\frac{Q_m \,r}{2\Xi} \\
{}^{\star}P_{\theta t}^{NLE}&= {}^{\star}P_{ \theta t}^{KN}+\xi\frac{a^2 Q_m\,\sin\theta\cos\theta}{2\Xi},
\end{split}
\end{equation} 
where the  KN fields $F_{\mu \nu}^{KN}$ and ${}^{\star}P_{\mu \nu}$ are given in Eqs. (\ref{Fmunu.KN}) and (\ref{Pmunu.KN}). Besides, the EM fields satisfy the alignments conditions Eqs. (\ref{AlignF}) and (\ref{AlignmP}).

This NLE generalization of the KN solution is new and we called it {\it the quartic vector potential} solution and its metric is given by Eq. (\ref{metric1}) with $\Delta_r$  given by
\begin{equation}\label{QDelta}
\Delta_r =r^2-2mr+ a^2 -\frac{\Lambda r^2}{3}(r^2+a^2)+(Q_e^2+Q_m^2)(1+\xi r^3),
\end{equation}
and the structural function $K(r)$ 

\begin{equation}
K(r)= (Q_e^2+Q_m^2) (1+\xi r^3),
\end{equation}
where $\xi$ is the nonlinear parameter. The other parameters are
the mass $m$, angular momentum $a$, electric and magnetic charges, $Q_e,~Q_m$ and the  cosmological constant $\Lambda$.
The Kerr-Newman solution is recovered by making $\xi=0$.
In the next section some features of this solution are explored.
\section{The quartic vector potential NLE-KN}

Since the quartic vector potential NLE generalization of the Kerr-Newman solution has not been reported elsewhere, as far as we know, we show that it represents a rotating black hole with a NLE field characterized by a NLE parameter $\xi$ whose introduction induces a de Sitter effect, with a cosmological horizon.

The metric function (\ref{QDelta}) is a fourth order polynomial that has at least one real positive root, therefore it represents a BH. To analyze the horizons determined by
$\Delta_{r}(r)=0$  we restrict to the case $\Lambda = 0$, that corresponds to a cubic equation in $r$; we also restrict to a vanishing magnetic charge, $Q_m=0$, but it can be recovered making $Q_e^2 \to Q_e^2 + Q_m^2$.

\subsection{Horizons and ergosphere without cosmological constant}
 
Let us consider the cubic equation (\ref{QDelta}) with $\Lambda=0$,
\begin{equation}\label{Delta.r.cub}
 \Delta_r=Q_e^2\xi r^3+r^2-2mr+Q_e^2+a^2=0.
\end{equation}
First we show that the equation always has at least one positive real root.  Let us notice that at $r=0,\quad \Delta_r=a^2+Q_e^2>0$. Furthermore, the dominant term is $Q_e^2\xi r^3$. In Section VI we show that $\xi < 0$; therefore, when $r\rightarrow \infty$ the function $\Delta_r\rightarrow -\infty$. Since the function changes sign in the interval $(0,\infty)$ it must cross the r-axis at least once and since the polynomial is cubic with real coefficients it has at least one positive real root; then the existence of an event horizon is guaranteed. 
\begin{figure}[ht]
\includegraphics[width=8cm,height=5cm]{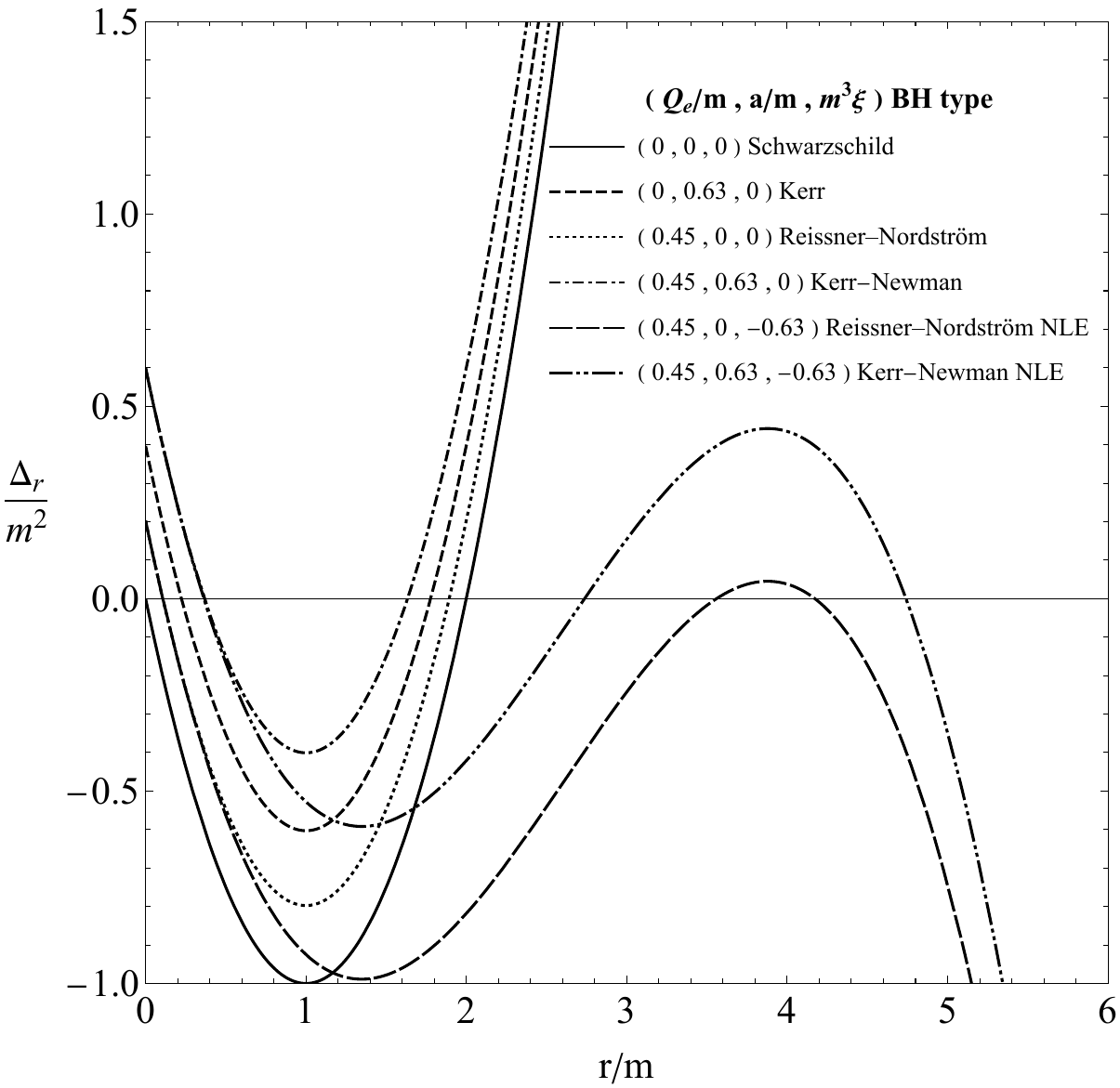}
\caption{\label{Delta.r} Graphics of the behavior of the metric function $\Delta_r (r) $ for the cases Schwarzschild, RN, Kerr, KN, NLE-RN and NLE-KN. } 
\end{figure} 

In this case, we will refer to the roots of $\Delta_r (r) = 0$ as the event horizon $(r_+)$, inner horizon $(r_-)$ and, a  third horizon induced by NLE,  cosmological horizon $(r_c)$; besides $r_-\leq r_+\leq r_c$. Fig. (\ref{Delta.r}) shows the behavior of $\Delta_r(r)$ for different sets of parameters $(m,a,Q_e,\xi)$. The relationships between the BH parameters and the roots in Eq. (\ref{Delta.r.cub}), by the Cardano-Vieta formulas,  are reduced to

\begin{align}
r_-&=\frac{2m(r_++r_c)-r_+r_c}{r_+ +r_c-2m},\label{r1.cub}\\
\xi  &= \frac{2m-r_+-r_c}{Q_e^2(r_+^2+r_+r_c+r_c^2)},\label{xi.cub}\\
a^2+Q_e^2&=\frac{r_+ r_c\left[2m(r_++r_c)-r_+r_c\right]}{r_+^2+r_+r_c+r_c^2}.\label{aQ.cub}
\end{align}

The cubic equation has three cases. The first one is  of roots of  multiplicity three, i.e. the three horizons overlap $(r_-= r_+ = r_c)$; this corresponds to $a^2+Q_e^2=\frac{4m^2}{3}$, $ \xi=-\frac{1}{6mQ_e^2}$ and the event horizon is at $r=2m$. 

The other two cases are roots of multiplicity two. One of them is the coincidence of the outer and cosmological horizon, $(0<r_- < r_+ = r_c)$, that can be analyzed using

\begin{equation}
 r_+=2m-(r_-) +\sqrt{{(r_-)}^2-2m (r_-)+4m^2}.
\end{equation}
The domain of this function is $0 < r_-<2m$ and in this interval the repeated roots are in the interval $(2m,4m)$. The case of coincidence of the inner and outer horizons, $(0<r_- = r_+ < r_c)$, can be analyzed using 
\begin{equation}
 r_+=2m-(r_c) +\sqrt{{(r_c)}^2-2m (r_c)+4m^2}.
\end{equation}
The domain of this function is $2m<r_c<\infty$ and the repeated roots are in the interval $(m,2m)$. For some values of the NLE parameter, $\xi$, there is not an event horizon but an inner horizon.

To classify the type of roots, we analyze the discriminant of the cubic equation,

\begin{align}\label{Delta.cub}
 \Delta_D&=-27(a^2+Q_e^2)^2 Q_e^4\left(\xi-\xi_+\right)\left(\xi-\xi_-\right),
\end{align}

where

\begin{equation}
    \xi_\pm=\frac{2\left[m(8m^2-9a^2-9Q_e^2)\pm(4m -3a^2-3Q_e^2)^\frac{3}{2}\right]}{27Q_e^2(a^2+Q_e^2)^2}.
\end{equation}

According to the sign of the discriminant there are the following cases: $\Delta_D=0$,  repeated roots; $\Delta_D>0$, three different real roots; and $\Delta_D<0$, one real root and two complex conjugated.

The case of repeated roots occurs when $\xi=\xi_\pm$ in (\ref{Delta.cub}). Taking into account that $\xi$ must be real, the electric charge $Q_e$ and the angular momentum $a$ are restricted to the interval $0<a^2+Q_e^2\leq \frac{4m^2 }{3}$. The function $\xi_+\in (0,\infty)$ for the values $0<a^2+Q_e^2< m^2$ and $\xi_+\in (-\infty,0]$ for the values $m^2\leq a^2+Q_e^2\leq \frac{4m^2}{3}$. On the other hand, the function $\xi_-\in (-\infty,0)$ for all values $0< a^2+Q_e^2\leq \frac{4m^2}{3}$. Furthermore, the inequality $\xi_-\leq\xi_+$ always holds. 

To solve the equation (\ref{Delta.r.cub}),    the transformation $r= z-\frac{1}{3Q_e^2\xi}$,  allows to depress the cubic equation to one without quadratic term,

\begin{equation}\label{cub.dep}
 Q_e^2\xi\left(z^3+\mathcal{M}z+\mathcal{N}\right)=0,
\end{equation}
where
\begin{eqnarray}\label{MN.cub}
\mathcal{M} & = & -\frac{1+6mQ_e^2 \xi}{3Q_e^4\xi^2},\nonumber\\
\mathcal{N} & = & \frac{2+18mQ_e^2 \xi+27(a^2+Q_e^2)Q_e^4\xi^2}{27Q_e^6\xi^3}.    
\end{eqnarray}

Next, with the change $z=2\sqrt{-\frac{\mathcal{M}}{3}}\cos\phi$ in (\ref{cub.dep}),  it reduces to  

\begin{equation}
    \cos 3\phi-\frac{3\mathcal{N}}{ 2\mathcal{M}}\sqrt{-\frac{3}{\mathcal{M}}}=0.
\end{equation}
Thus, the roots $r_k$ of (\ref{Delta.r.cub}) are

\begin{equation}\label{root.cub}
\begin{split}
    r_{k+1}&=2\sqrt{-\frac{\mathcal{M}}{3}}\cos \left[\frac{1}{3}\arccos{\left(\frac{3\mathcal{N}}{ 2\mathcal{M}}\sqrt{-\frac{3}{\mathcal{M}}}\right)+ \frac{2\pi k}{3} }\right]\\
    &-\frac{1}{3Q_e^2\xi},
\end{split}
\end{equation}
where $k=0,1,2.$
The previous solution demands that:
\begin{align}
 0&<-\frac{3}{\mathcal{M}},\\
 -1&\leq \frac{3\mathcal{N}}{2\mathcal{M}}\sqrt{-\frac{3}{\mathcal{M}}}\leq 1.
\end{align}
These restrictions reduce the ranges of the parameters as follows: $a\in \left[0,\frac{2m}{\sqrt{3}}\right)$, $Q_e\in \left(0,\frac{2m}{\sqrt{3}}\right]$ and $\xi\in \left(-\infty,0\right)$. 
Analyzing the derivatives of the function $\Delta_{r}(r)$, we can find a function for $\xi$,
\begin{equation}\label{Delta.prime}
\Delta_{r}^{\prime}= 3Q_e^2\xi r^2+2r-2m =0,\longleftrightarrow  Q_e^2\xi=\frac{2m}{3r^2}-\frac{2}{3r}.   
\end{equation}
For $m\leq r$, we have that $-\frac{1}{6m}\leq Q_e^2 \xi \leq 0$. By substituting the term $Q_e^2 \xi$ into $\Delta_r$ we can find that its roots are given by the equation

\begin{equation}\label{Delta.cub.crit}
\Delta_r=\frac{r^2}{3}-\frac{4mr}{3}+a^2+Q_e^2,
\end{equation}
whose internal horizons $(hr_-)$ and external horizons $(hr_+)$ are

\begin{equation}
     hr_\pm=2m\pm\sqrt{4m^2-3(a^2+Q_e^2)},
\end{equation}
This expression for the horizon is only valid when $a^2+Q_e^2 \leq \left(\frac{2m}{\sqrt{3}}\right)^2 \approx (1.1547m)^2$. 
Additionally, the ergosphere is obtained from the roots of  $g_{tt}=0$. In this case, $g_{tt}= (a^2\sin^2\theta-\Delta_r)$;
so, using the previously roots with $a^2 \to a^2 \cos^2 \theta,$ the radius of the ergosphere is given by
\begin{equation} 
r_{ergo}=2m\pm\sqrt{4m^2-3\left(a^2\cos^2\theta+Q_e^2\right)}
\end{equation}

\begin{figure}[ht]
\includegraphics[width=8cm,height=5cm]{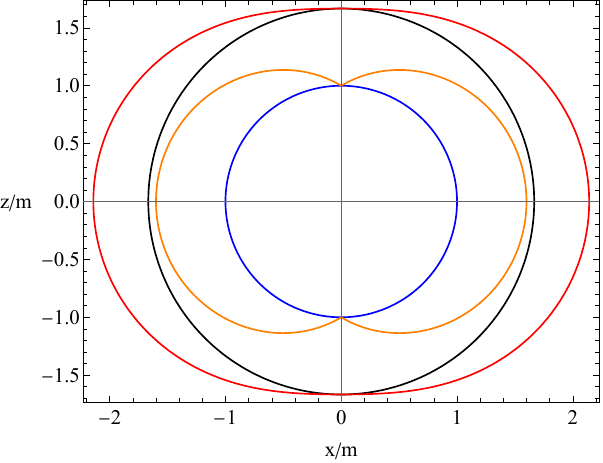}
\caption{\label{Ergo} The event horizon for KN (blue) and KN NLE (black), as well as the ergosphere of KN (orange) and the one of KN NLE (red) are displayed. The BH parameters are set to $Q_e = 0.8m$, $a = 0.6m$ and $\xi = -0.14/m^3$.} 
\end{figure}

On the other hand $\Delta_r^{\prime}=0$ (\ref{Delta.prime}) has two roots,

\begin{equation}
r_{1,2}=  \frac{1 \pm \sqrt{1+ 6m Q_e^2 \xi}}{-3 Q_e^2  \xi},    
\end{equation}
considering that $\xi$ is negative, the two roots are positive,
and $r_{1} > r_{2}$. Evaluating the second derivative,

\begin{eqnarray}
\Delta_{r}^{\prime \prime} (r_{1} ) & = & -2 \sqrt{1+ 6mQ_e^2 \xi} < 0, \nonumber\\
\Delta_{r}^{\prime \prime} (r_{2} ) & = & +2 \sqrt{1+ 6mQ_e^2\xi} > 0,
\end{eqnarray}

then the function $\Delta_{r}$ has a minimum at $r_{2} $ and a maximum at $r_{1} $ with  $r_{2} < r_{1}$, such that  there are  three horizons, being the third horizon an  effect of the nonlinear electromagnetic field. For some particular values of the parameters $(a, Q_e, m, \xi)$ the plots of ergosphere and event horizon are shown in Fig. \ref{Ergo}.

\subsection{The static limit of the NLE Quartic vector potential solution}

In the limit of vanishing rotation we obtain a NLE generalization of the Reissner-Norsdtrom solution with cosmological constant, given by the static metric,

\begin{eqnarray}
ds^2 & = & -f(r) dt^2+ f(r)^{-1}dr^2 + r^2(\sin^2 \theta d\theta + d \phi^2),\nonumber\\
f(r) & = & 1 - \frac{2m}{r} + \frac{Q_e^2 + Q_m^2}{r^2}(1+ \xi r^3)+ \frac{\Lambda}{3}r^2,
\end{eqnarray}
characterized by the electromagnetic fields

\begin{eqnarray}
{\Xi} F_{rt}&=&{ \frac{Q_e}{2r^2} \left( \xi r^3 -2 \right)}, \nonumber\\
{\Xi} {}^*P_{rt}&=& {\frac{Q_m}{2r^2} \left( \xi r^3 -2 \right)}, 
\end{eqnarray}
This NLE-RN also presents one cosmological horizon and other interesting features, however, the analysis of this solution is beyond the scope of this paper.

\section{The nonlinear electromagnetic energy momentum tensor}

To asses if the NLE energy-momentum tensor $T_{\mu \nu}$  is physically reasonable, by this meaning that the local energy density measured by any observer appears non-negative  as well as  the local energy flow vector be non-spacelike, we check the energy conditions   associated to the NLE generalization of the KN. 
To this end we project  $T_{\mu \nu}$ onto the orthonormal tetrad, $\{\bf {E^a},a=1,\ldots,4\}$ associated to the metric (\ref{metric1}),  that is given by:

\begin{eqnarray}\label{orthob}
{\bf E^1} & = &\sqrt{\frac{\Sigma}{\Delta_\theta}} d\theta, \quad {\bf E^3}=\sqrt{\frac{\Sigma}{\Delta_r}} dr,\nonumber\\
{\bf E^2} & = & \frac{\sin\theta}{\Xi}\sqrt{\frac{\Delta_\theta}{\Sigma}}\left[adt-\left(r^2+a^2\right)d\phi\right],\nonumber\\
{\bf E^4} & = &\frac{1}{\Xi}\sqrt{\frac{\Delta_r}{\Sigma}}\left(dt-a\sin^2\theta d\phi\right).
\end{eqnarray}

Then projecting  $T_{\mu \nu}$ onto the orthonormal basis its canonical form is obtained,
\begin{equation}
\label{Orto.Tmunu}
OT^{ab}=
\begin{pmatrix}
\frac{T_{\theta \theta}}{g_{\theta \theta}}  & 0 & 0 & 0\\
0& \frac{T_{\theta \theta}}{g_{\theta \theta}} & 0 & 0 \\
0 & 0& \frac{T_{r r}}{g_{r r}}  & 0 \\
0 & 0& 0&  -\frac{T_{r r}}{g_{r r}} 
\end{pmatrix}
=\begin{pmatrix}
        p_1&0&0&0\\
        0&p_2&0&0\\
        0&0&p_3&0\\
        0&0&0&\mu
    \end{pmatrix}.
\end{equation}
Where $p_1,~p_2,~p_3$ represent the principal {\it pressures} in the three spacelike directions $\mathbf{E^\alpha}(\alpha=1,2,3)$. The eigenvalue $\mu$ represents the energy density as measured by and observer whose world-line at point $p$ has a unit tangent vector $\mathbf{E^4}$. Note that the $OT^{ab}$ tensor has the canonical form type I (\cite{Hawking}). 
In terms of the EM potentials and the metric function $K(r)$ the $OT^{ab}$ components are

\begin{eqnarray}\label{Trr}
- \mu & = & p_3 = \frac{T_{rr}}{g_{rr}}= 
-\sqrt{F^2+G^2} \frac{\mathcal{L}_F}{4\pi}+\frac{K^{\prime \prime}(r)}{32\pi \Sigma} \nonumber\\
& = &
\frac{\Xi^2}{8 \pi a\sin\theta}\left(\frac{\partial A_t}{\partial r}\frac{\partial {}^{\star}P_t}{\partial \theta}-\frac{\partial A_t}{\partial \theta}\frac{\partial {}^{\star}P_t}{\partial r}\right)+\frac{K^{\prime \prime}(r)}{32\pi \Sigma} \nonumber\\
\end{eqnarray}

\begin{eqnarray}\label{T.theta.theta}
p_1 & = & p_2=\frac{ T_{\theta\theta}}{g_{\theta\theta}}=
\sqrt{F^2+G^2} \frac{\mathcal{L}_F}{4\pi}+\frac{K^{\prime \prime}(r)}{32\pi \Sigma} \nonumber\\
&=&
\frac{-\Xi^2}{8 \pi a\sin\theta}\left(\frac{\partial A_t}{\partial r}\frac{\partial {}^{\star}P_t}{\partial \theta}-\frac{\partial A_t}{\partial \theta}\frac{\partial {}^{\star}P_t}{\partial r}\right)+\frac{K^{\prime \prime}(r)}{32\pi \Sigma},\nonumber\\
\end{eqnarray}
\subsection{Energy conditions}
In what follows we determine the conditions on the $OT^{ab}$ components in order to fulfil the weak, dominant and strong energy conditions,  that are summarized in Table \ref{table.EC}.

\begin{table}[h!]
\centering
\begin{tabular}{|c | c c |c|} 
 \hline
 Energy Cond. & $OT_{ab}$ components & & Inequalities\\ 
 \hline
 Weak & $T_{ab}u^a u^b\geq 0$ & &  $\mu\geq0$, $\mu+p_\alpha\geq 0$ \\
 \hline
 Dominant & $T_{ab}u^a u^b\geq 0$, $T^{ab}u_b \leq 0$ & & $\mu\geq 0$, $\mu\geq |p_\alpha|$\\
 \hline
\multirow{2}{*}{Strong}&\multirow{2}{*}{$(T_{ab}-\frac{1}{2}Tg_{ab})u^a u^b\geq 0$} && $\mu+\sum_{\alpha=1}^{3} p_\alpha\geq 0$,\\
& &    &$\mu+p_\alpha\geq 0$\\
\hline
\end{tabular}
\caption{Energy conditions. The inequalities that the stress-energy tensor 
$T_{ab}$ should satisfy to fulfill the energy conditions; $u^{a}$  is a timelike vector; $\mu = - OT_{rr}/g_{rr}= -p_3$ and $p_1=p_2= OT_{\theta \theta}/g_{\theta \theta}$.}
\label{table.EC}
\end{table}

\subsubsection{ Weak Energy Condition (WEC)}
If the energy momentum tensor is of type I, the Weak Energy Condition (WEC)\cite{Hawking} holds if $\mu\geq0$. Additionally, the energy density should not be exceeded by any {\it pressure}, such that $\mu+p_{\alpha} \geq 0,\quad \alpha= 1, 2, 3$.
In the case of  the quartic vector potential NLE-KN solution WEC amounts to
the following conditions,

\begin{equation}\label{WEC.1}
\mu=\frac{\tilde{Q}^2}{8\pi\Sigma^2}\left(1-2\xi r^3\right) \ge 0, \quad \mu+p_3=0.\nonumber
\end{equation}

\begin{align}\label{WEC.2}
 \mu+p_{1,2}=\frac{\tilde{Q}^2}{8\pi \Sigma^2}\left(2+3\xi ra^2\cos^2\theta-\xi r^3\right)\geq 0, & & 
\end{align}
where $\tilde{Q}^2=(Q_e^2+Q_m^2)$
The conditions (\ref{WEC.1}) are fulfilled if the NLE parameter is less than zero, $\xi <0$. If $\xi=0$ the KN solution is recovered.
If $\xi>0$ Eqs. (\ref{WEC.1})   reduce to $0\leq \xi r^3 \leq \frac{1}{2}$, such that as $r$ approaches infinity, the only value that keeps within this range is $\xi=0$.

\subsubsection{ Dominant Energy Condition (DEC)}
The Dominant Energy Condition (DEC)\cite{Hawking} holds if
$\mu\geq 0$, and $-\mu- p_\alpha\leq 0\leq \mu-p_\alpha$. This leads to the following inequalities for the NLE-KN

\begin{align}\label{DEC.1}
-(\mu+p_{1,2})\leq & 0\leq \mu-p_{1,2}=-\frac{T}{2}\nonumber\\
    \frac{\tilde{Q}^2}{8\pi \Sigma^2}\left[\xi r^3-3\xi r a^2 \cos^2\theta-2\right]\leq& 0\leq -\frac{3 \tilde{Q}^2 \xi r}{8\pi \Sigma}.
\end{align}
The case for $p_3$ reduces to $0\leq 2\mu$.  If $\xi\leq0$ then (\ref{DEC.1}) are automatically satisfied.
\subsubsection{ Strong Energy Condition (SEC)}
The SEC guarantees that matter is attractive, causing geodesics to converge. In terms of the $OT_{ab}$ components SEC amounts to $\mu+p_1+p_2+p_3 \geq 0$.  
For the quartic vector potential NLE-KN solution this reduces to

\begin{equation}
0 \leq \frac{\tilde{Q}^2}{4\pi \Sigma^2}\left(1+3\xi ra^2\cos^2\theta+\xi r^3\right),
\end{equation}
the above inequality is satisfied for $\xi=0$ (KN case); however if $\xi<0$ the equation is violated when $r\rightarrow\infty$ by a large negative pressure. The case $\xi>0$ causes WEC and DEC to be violated, therefore the most sensible choice  for the quartic potential generalization of the KN solution is $\xi<0$. 
Note that even SEC is violated by the introduction of the NLE parameter $\xi$, this does not make less meaningful the NLE-KN solution since SEC can be violated by certain forms of matter such as a massive scalar field and quantum fields can generically violate any of the energy conditions \cite{Carrol2004}. 
\subsection{Weyl conformal symmetry}

From the geometric part of Einstein's equations, that is, from the left hand  side of 
$G_{\mu\nu}+\Lambda g_{\mu\nu}=8\pi T_{\mu\nu}$, the trace of the energy-momentum tensor can be determined as

\begin{align}
8 \pi \tensor{T}{^\mu_\mu} & = \frac{1}{\Sigma} \left[ \Delta_{r}{}^{\prime \prime} - 2 + \left(\frac{2}{3}a^2+4r^2 \right)\Lambda \right] \nonumber\\
&=\frac{K^{\prime\prime}(r)}{\Sigma}.
\end{align}
Therefore for the quartic vector potential NLE-KN solution, since $K^{\prime\prime}(r) \ne 0$,
 then the conformal invaruance is broken by the NLE field.

The Maxwell stress-tensor is trace-free; in agreement that  for the Kerr-Newman solution 
\begin{eqnarray}
\Delta_{r} &=& r^2-2mr+ a^2 + Q_e^2 - \frac{\Lambda}{3}r^2(r^2+a^2),\nonumber\\
\Delta_{r}^{\prime \prime} &=&  2 - \left(\frac{2}{3}a^2+4r^2 \right)\Lambda, 
\end{eqnarray}
with $K(r)= Q_e^2$ then $K^{\prime \prime}(r)=0$ and
the trace-free condition is fulfilled.  Moreover for any NLE that preserves conformal invariance in a Kerr-like geometry, the condition to be trace-free demands $K(r)^{\prime \prime} =0.$

For instance the ModMax generalization consists in transforming 
$Q_e^2 \mapsto e^{-\gamma} Q_e^2$, then preserving the $T_{\mu \nu}$ traceless.
\subsection{Lagrangian for the quartic vector potential solution}

The Lagrangian for the Kerr-Newman metric in terms of the coordinates is given by

\begin{equation}
    \mathcal{L}_{KN}=\frac{h_1(r,\theta) h_2(r,\theta)}{2(r^2+a^2\cos^2\theta)^4},
\end{equation}
where the functions $h_1(r,\theta)$ and $h_2(r,\theta)$ are
\begin{align}
    h_1(r,\theta)&=r^2(Q_e-Q_m)+2ra\cos\theta(Q_e+Q_m)\nonumber\\
    &-a^2\cos^2\theta(Q_e-Q_m),\nonumber\\
    h_2(r,\theta)&=r^2(Q_e+Q_m)-2ra\cos\theta(Q_e-Q_m)\nonumber\\
    &-a^2\cos^2\theta(Q_e+Q_m).   
\end{align}

While from the matter content of the stress-energy tensor,  

$4\pi T_{\mu\nu}=\mathcal{L} g_{\mu\nu}-\tensor{F}{_\mu^\alpha}P_{\nu\alpha}$, 

the expression for the Lagrangian $\mathcal{L}$ is found as,

\begin{eqnarray}
\mathcal{L}& = &-\frac{\Xi^2}{2a\sin\theta}\left(\frac{\partial A_t}{\partial r}\frac{\partial {}^{\star}P_t}{\partial \theta}+\frac{\partial A_t}{\partial \theta}\frac{\partial {}^{\star}P_t}{\partial r}\right)+\frac{K^{\prime\prime}(r)}{8\Sigma}.
\end{eqnarray}
Then for the quartic vector potential NLE-KN,
the  Lagrangian $\mathcal{L}_{NLE}(r, \theta)$ is given by
\begin{eqnarray}
\mathcal{L}_{NLE}(F,G) & = &\mathcal{L}_{KN}+\xi\frac{\left[(Q_e^2+2Q_m^2)r-Q_e Q_m\,a\cos\theta\right]}{2(r^2+a^2\cos^2\theta)}\nonumber\\
&& - \xi^2\frac{Q_e Q_m\,ra\cos\theta }{4}.
\end{eqnarray}
And we  recover $\mathcal{L}_{KN}$ making $\xi=0$.

\section{Conclusions}\label{sect6}

We present in detail the method to find nonlinear electromangetic (NLE) solutions in a Kerr-like metric, that previously was introduced in \cite{AGarcia_Annals2022}.  
We then examine the general form of the electromagnetic potentials to determine exact solutions of the coupled NLE-Einstein equations in a Kerr-like geometry; and it was found that there are only two possible NLE generalizations;  one of them was already presented in \cite{AGarcia_Annals2022}. The second case, that we called {\it quartic vector potential solution}, is new and corresponds  to a NLE generalization of the Kerr-Newman black hole characterized by  the introduction of a NLE parameter that induces a third horizon, resembling a cosmological horizon. The electromagnetic fields,  horizons and ergosphere of the NLE quartic vector potential generalization of the Kerr-Newman (KN) solution are presented. The static limit corresponds to a NLE generalization of the Reissner-Nordstrom BH with cosmological constant. 

Additionally, we determine the canonical form of the NLE stress-energy tensor and set up the inequalities to fulfill the physically reasonable energy conditions. Imposing the energy conditions to the found solution we find as a constraint that the nonlinear parameter $\xi$ should be negative. The trace of the NLE stress-energy tensor  does not vanish, then the NLE field breaks conformal invariance.
Moreover, the expression of the NLE Lagrangian  $\mathcal{L}_{NLE}(r, \theta)$ is presented, its form consists of two terms, the KN Lagrangian plus a term derived from the NLE contribution.

The nature and precise interpretation of the NLE field deserves further investigation, for instance if there are electromagnetic multipoles associated to the NLE fields; we leave this for a future work.



{\bf Acknowledgements}

NB acknowledges partial support by CONAHCYT-Mexico 
project CBF-2023-2024-811. The work of OG has been sponsored by CONAHCYT-Mexico through the Ph. D.  scholarship No. 815804. OG also acknowledges hospitality by Profr. C. Laemmerzahl at the  Universität Bremen where part of this work was done.


\end{document}